\documentclass[journal=jacsat,manuscript=communication,etalmode=truncate,keywords=true]{achemso}
\setkeys{acs}{articletitle=true,maxauthors=10,etalmode=truncate,keywords=true}
\usepackage{amsmath,amssymb,units,txfonts,upgreek}
\usepackage{dblfloatfix}
\usepackage{graphicx}
\usepackage{helvet}
\usepackage[usenames,dvipsnames]{color}
\definecolor{StyleColor}{RGB}{34,80,169} 
\definecolor{abstractcolor}{RGB}{255,243,201} 
\usepackage[bookmarks=true,colorlinks=true,urlcolor=StyleColor,linkcolor=StyleColor,citecolor=StyleColor]{hyperref} 
\usepackage{colortbl,cuted,bm}
\usepackage{multirow}
\usepackage{xr}

\newcommand{\old}[1]{}
\usepackage{xfrac}
\DeclareMathOperator{\Imag}{Im}

\makeatletter\newenvironment{abstractbox}{%
   \begin{lrbox}{\@tempboxa}\begin{minipage}{0.988\textwidth}}{\end{minipage}\end{lrbox}%
   \colorbox{abstractcolor}{\usebox{\@tempboxa}}
}\makeatother

\renewcommand*\authorfont{\flushleft}
\renewcommand*\affilfont{\mdseries\flushleft\textrm}
\renewcommand*\titlesize{\Large}
\renewcommand*\titlefont{\flushleft\sf\bfseries}
\def\refname{\sffamily\color{StyleColor}
\large$\blacksquare$\normalsize\ 
REFERENCES}
\usepackage{titlesec}
\titleformat{\section}{\bfseries\sffamily\color{StyleColor}}{\thesection.~}{0pt}{}
\titleformat{\subsection}[runin]{\bfseries\sffamily\normalsize}{\indent\thesubsection.~}{0pt}{}[.]
\titlespacing{\subsection}{0pt}{0pt}{*1}
\titleformat{\subsubsection}{\bfseries\sffamily\normalsize}{\thethesubsection.~}{0pt}{}
\titlespacing{\subsubsection}{0pt}{0pt}{*0}

\title{Photoinduced Absorption within Single-Walled Carbon Nanotube Systems}
\author{Livia No\"{e}mi Glanzmann}
\affiliation[UPV/EHU]{\newline\footnotemark[2]{\ } Nano-Bio Spectroscopy Group and ETSF Scientific Development Center, Departamento de F{\'{\i}}sica de Materiales, Centro de F\'{\i}sica de Materiales CSIC-UPV/EHU-MPC 
 Universidad del Pa{\'{\i}}s Vasco UPV/EHU and DIPC, E-20018 San Sebasti\'{a}n, Spain}
\author{Duncan John Mowbray}
\email{duncan.mowbray@gmail.com}
\affiliation[UPV/EHU]{\newline\footnotemark[2]{\ } Nano-Bio Spectroscopy Group and ETSF Scientific Development Center, Departamento de F{\'{\i}}sica de Materiales, Centro de F\'{\i}sica de Materiales CSIC-UPV/EHU-MPC 
 Universidad del Pa{\'{\i}}s Vasco UPV/EHU and DIPC, E-20018 San Sebasti\'{a}n, Spain}
\author{Diana Gisell Figueroa del Valle}
\affiliation[IIT]{\newline\footnotemark[3]{\ } Center for Nano Science and Technology, Istituto Italiano di Tecnologia, Via Pascoli 70/3, I-20133 Milano, Italy}
\alsoaffiliation[PM]{\newline\footnotemark[5]{\ } Politecnico di Milano, Dipartmento di Fisica, P.zza L. Da Vinci 32, I-20133 Milano, Italy}
\author{Francesco Scotognella}
\affiliation[IIT]{\newline\footnotemark[3]{\ } Center for Nano Science and Technology, Istituto Italiano di Tecnologia, Via Pascoli 70/3, I-20133 Milano, Italy}
\alsoaffiliation[PM]{\newline\footnotemark[5]{\ } Politecnico di Milano, Dipartmento di Fisica, P.zza L. Da Vinci 32, I-20133 Milano, Italy}
\author{Guglielmo Lanzani}
\affiliation[IIT]{\newline\footnotemark[3]{\ } Center for Nano Science and Technology, Istituto Italiano di Tecnologia, Via Pascoli 70/3, I-20133 Milano, Italy}
\alsoaffiliation[PM]{\newline\footnotemark[5]{\ } Politecnico di Milano, Dipartmento di Fisica, P.zza L. Da Vinci 32, I-20133 Milano, Italy}
\author{Angel Rubio}
\affiliation[UPV/EHU]{\newline\footnotemark[2]{\ } Nano-Bio Spectroscopy Group and ETSF Scientific Development Center, Departamento de F{\'{\i}}sica de Materiales, Centro de F\'{\i}sica de Materiales CSIC-UPV/EHU-MPC 
 Universidad del Pa{\'{\i}}s Vasco UPV/EHU and DIPC, E-20018 San Sebasti\'{a}n, Spain}
\alsoaffiliation[MaxPlanck]{\newline\footnotemark[4]{\ } Max Planck Institute for the Structure and Dynamics of Matter, Luruper Chaussee 149, D-22761 Hamburg, Germany}

\begin{document}
\maketitle

\begin{strip}
\vspace{-1.cm}

\noindent{\color{StyleColor}{\rule{\textwidth}{0.5pt}}}
\begin{abstractbox}
\begin{tabular*}{17cm}{b{11.5cm}@{\ }r}
\noindent\textbf{\color{StyleColor}{ABSTRACT:}}
We study the photoabsorption properties of photoactive bulk polymer/ fullerene/nanotube heterojunctions in the near-infrared region. By combining pump-probe spectroscopy and linear response time-dependent density functional theory within the random phase approximation (TDDFT-RPA) we elucidate the excited state dynamics of the $E_{11}$ transition within (6,5) and (7,5) single-walled carbon nanotubes (SWNTs) and combined with poly(3-hexylthiophene-2,5-diyl) (P3HT) and [6,6]-phenyl-C$_{61}$-butyric acid methyl ester (PCBM) in P3HT/PCBM/SWNT blended samples. We find the presence of a photoinduced absorption (PA) peak is related mainly to the width of the photobleach (PB) peak and the charge carrier density of the SWNT system. For mixed SWNT samples, the PB peak is too broad to observe the PA peak, whereas within  P3HT/PCBM/SWNT blended samples P3HT acts as a hole acceptor, narrowing the PB peak by exciton delocalization, which reveals a PA peak. Our results suggest that the PA peak originates from a widening 
of the band gap in the presence of excited  electrons and holes. These results have important implications for the development of new organic photovoltaic heterojunctions including~SWNTs. &
\includegraphics[width=5.7cm]{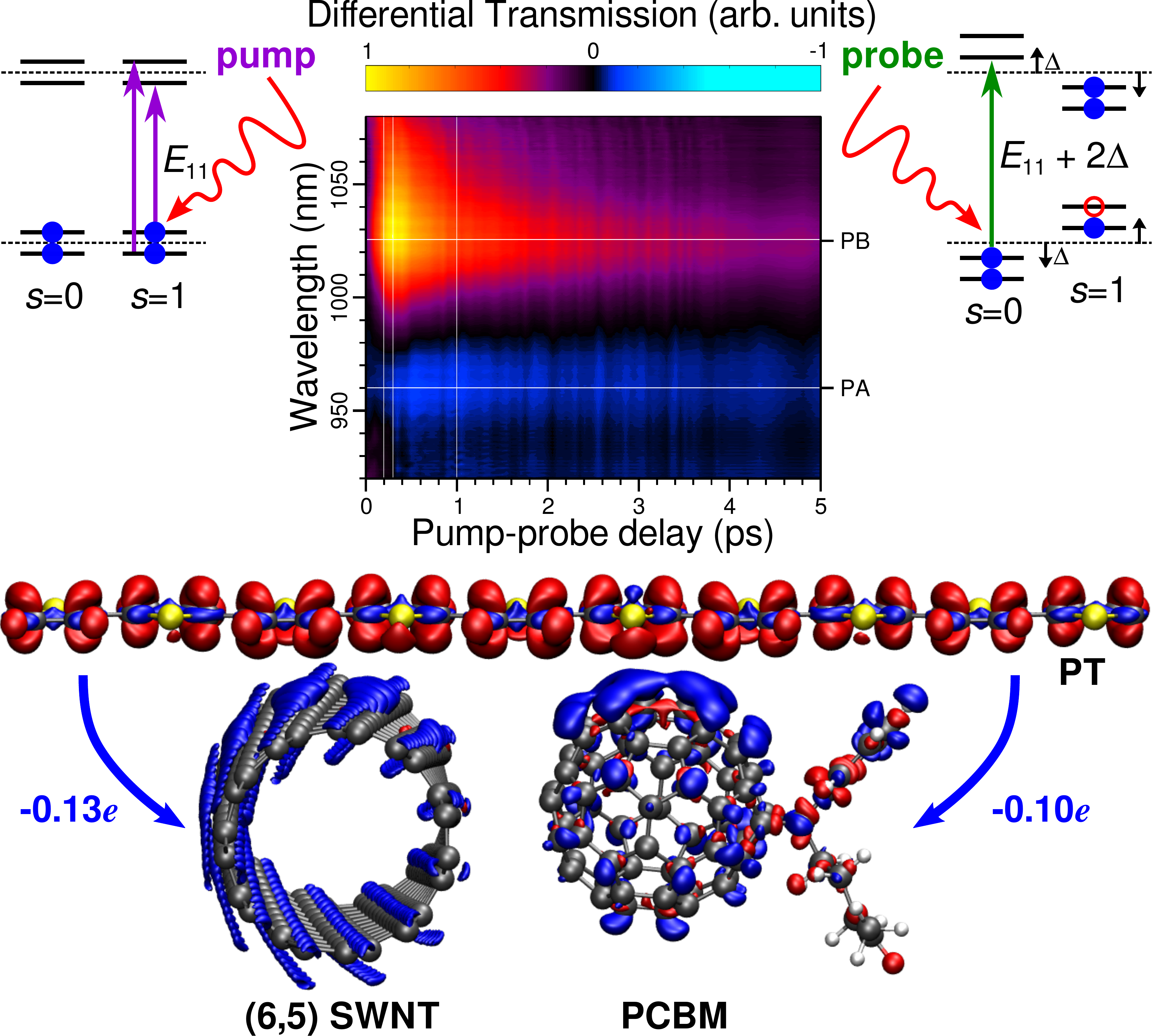}\\
\end{tabular*}
\end{abstractbox}
\noindent{\color{StyleColor}{\rule{\textwidth}{0.5pt}}}
\end{strip}

\section{INTRODUCTION}\label{Sect:Introduction}

Single-walled carbon nanotubes (SWNTs) \cite{Iijima,Harris,Dresselhaus}, with their high charge carrier densities \cite{currentdensitySWNT} and mobilities \cite{mobility, holemobility, electronmobility},  show promise as electron transporting materials for photoactive polymers \cite{Ferguson} within organic photovoltaic devices (OPVs).  In fact, a significant enhancement of photocurrent was achieved by inserting SWNTs into the bulk heterojunction (BHJ) of an OPV \cite{Kymakis02,Kymakis03,HertelPumpProbeSWNT+PFO-BPy}.  OPVs with SWNTs can be further improved by enriching the content of semiconducting tubes\cite{Holt}, thereby reducing electron-hole trapping\cite{Strano}.  SWNTs have even been used directly as photoreceptors within BHJs\cite{BindlJPCL, Bindl-SWNT}.  However, the power conversion efficiencies (PCEs) of SWNT OPVs still lag behind the best BHJ OPVs\cite{Loi}.  The optimization of SWNT OPVs requires a combined theoretical and experimental approach to understand the photoelectric processes within SWNT composites. Only in so doing can we take full advantage of the outstanding physical properties of SWNTs  within BJH OPVs.

The elementary photoexcitations in SWNTs are singlet excitons with a large binding energy of about 400 meV. \cite{AndoJPSJ1997,AvourisPRL2004,SpataruPRL2004,HeinzScience2005,LienauPRB2005,SpataruPRL2010} Pump-probe spectroscopy is a powerful tool for studying optical properties and exciton dynamics over the femtosecond to microsecond time domain. As an example, the size of the exciton and its diffusion in the incoherent regime have been experimentally estimated for (6,5) SWNTs\cite{larry-exlength} using the phase space filling model.\cite{GreeneScience1990} Also, the energy of relaxation from higher exciton states to the first exciton $E_{11}$ transition has been measured, yielding a very short decay time of 50~fs.  This is due to the peculiarity of a one-dimensional system, such as a SWNT, and the strong coupling to highly energetic optical phonons.\cite{LanzaniPRL2005} The strong electron-phonon coupling is evident in the large Raman cross-section, extensively studied with standard continuous wave techniques. The most intense Raman active modes, radial breathing modes and G modes, have been detected in transient coherent Raman experiments.\cite{BishopNatPhys2006,LanzaniPRL2009}

Unfortunately, there are other photoexcited species, resulting from highly nonlinear phenomena, that might occur in SWNTs, besides singlet excitons. For this reason, the interpretation of transient spectra for SWNTs becomes difficult, and this is due to contributions of such species, e.g., triplets \cite{HertelPumpProbeSWNT+PFO-BPy}, biexcitons\cite{Styers-Barnett,width-theory}, trions\cite{trion,TobiasACSNano2015}, and charge-carrier photogeneration\cite{LanzaniJPCC2013,sr_Lanzani}. Previous transient absorption measurements\cite{Zhuetal, Zhu} have found a photoinduced absorption (PA) peak at $\sim 950$~nm for predominantly (6,5) SWNT samples when pumped in the visible (VIS) region.   The structure of this PA feature provides direct information about the photoexcitation processes within  the SWNT.  It is the structure and origin of this PA peak, and how it depends on the SWNT system, which we will focus upon herein.

In this work, we employ state-of-the-art pump-probe transient absorption spectroscopy to measure the excited state dynamics  of a (6,5) and (7,5) SWNT mixture, and combined with poly(3-hexylthiophene-2,5-diyl) (P3HT) and [6,6]-phenyl-C$_{61}$-butyric acid methyl ester (PCBM) in P3HT/PCBM/SWNT blended samples in the near-infrared (IR).  We then employ linear response time dependent (TD) density functional theory (DFT) in frequency-reciprocal space within the random phase approximation (RPA) to model the measured transient spectra for (6,5) and (7,5) SWNTs and blended P3HT/PCBM/SWNT samples.  

We begin by providing details of the sample fabrication in section~\ref{SampleFabrication}, optical characterization of the sample in section~\ref{OpticalCharacterization}, the theoretical methods employed in section~\ref{TheoreticalMethods}, and computational parameters used in section~\ref{ComputationalDetails}. In section~\ref{TransientAbsorptionMeasurements} we provide a detailed comparison of the absorbance and differential transmission spectra obtained for our SWNT and blended P3HT/PCBM/SWNT samples with those available in the literature in the near-IR region as a function of the pump-probe delay.  We show in section~\ref{InterpretingTransientAbsorptionasaDifferenceSpectra} how the differential transmission spectra may be interpreted in terms of a difference in absorption spectra. After briefly justifying our method for modeling optically excited states in section~\ref{ModelingOpticallyExcitedStates}, we employ TDDFT-RPA calculations of (6,5) SWNTs, (7,5) SWNTs, and a combined PT/PCBM/(6,5) SWNT system in section~\ref{TDDFTSimulatedSpectra} to simulate the measured differential transmission spectra, and explain the dependence of the observed PA peak on charge carrier density in section~\ref{InfluenceofCharging} and the photobleach (PB) peak width in section~\ref{InfluenceofPeakWidth}. This is followed by concluding remarks. A derivation of the TDDFT-RPA formalism, and the influence of local field effects on the SWNT spectra are provided in Appendix~\ref{LCAOTDDFTRPA}. Atomic units ($\hslash=m_e=e=a_0=1$) have been used throughout unless stated otherwise.

\section{METHODOLOGY}\label{Sect:Methodology}

\subsection{Sample Fabrication}\label{SampleFabrication}

The glass substrates were precleaned with acetone and isopropanol and dried under a flow of dry nitrogen. For preparation of the samples, regular P3HT poly(3-hexylthiophene-2,5-diyl) and PCBM ([6,6]-phenyl-C$_{61}$-butyric acid methyl ester) were dissolved in ortho-dichlorobenzene (ODCB) at a 1:1 ratio. We employed 704148-SWNTs produced using the CoMoCAT$^{\text{\textregistered}}$ catalytic chemical vapor deposition (CVD) method. The SWNTs were also dispersed in ODCB  and then sonicated for 1 hour. No debundling or removal of metallic SWNTs was performed on the sample, which was most likely aggregated. The resulting solution was then spin-coated on top of the glass substrates to obtain the SWNT sample. In the case of the P3HT/PCBM/SWNT sample, the dispersed SWNTs were added to the P3HT/PCBM solution at a 1:1:1 ratio and then sonicated for 1 hour. The solution was then spin-coated on top of the glass substrates at 1000 rpm for 2 minutes. All the materials were bought from Sigma-Aldrich.  

\subsection{Optical Characterization}\label{OpticalCharacterization}
The ground state absorption spectra were collected with a PerkinElmer spectrophotometer (Lambda 1050 WB InGaAs 3D WB Detection Module). The laser system employed for ultrafast transient absorption was based on a Ti-Sapphire chirp pulse amplified source; with a maximum output energy of about 800 $\upmu$J, 1~kHz repetition rate, central wavelength of 780~nm and pulse duration of about 180 fs. Excitation pulses at 590 and 900~nm  were generated by noncollinear optical parametric amplification in a $\upbeta$-barium borate (BBO) crystal, with a pulse duration of around 100~fs. Pump pulses were focused in a 200 $\upmu$m diameter spot. Probing was achieved in the visible and near IR region by using white light generated using a thin sapphire plate. Chirp-free transient transmission spectra were collected by using a fast optical multichannel analyzer (OMA) with a dechirping algorithm. The measured quantity is the differential transmission, $\Delta T = T(t) - T(t=0)$. Once normalized, the differential transmission $\Delta T/T$ may be directly compared with the change in absorbance $\Delta\Imag[\varepsilon] = \Imag[\varepsilon(t=0)] - \Imag[\varepsilon(t)]$. The excitation energy has been set to 11 nJ when pumping at 590~nm and then increased to 200 nJ when pumping at 900~nm, i.e., selectively pumping the SWNTs. All measurements were performed in air at room temperature.

\subsection{Theoretical Methods}\label{TheoreticalMethods}

To model differential transmission measurements, we use the difference between the optical absorption of the system in the ground and excited states.  We model the system in the excited state within DFT by fixing the total magnetic moment $\mu$, and through the addition of charge $Q$ to the system. Specifically, we use the singlet ($S = 0$) to model the ground state, the triplet ($S = 1$) to model a single exciton, the quintet ($S = 2$) to model a pair of excitons, and the quartet ($S = 3/2$) with an additional charge $Q = -e$ to model a negative trion, i.e., a pair of excited electrons and a single hole.  

The optical absorption spectra are obtained via linear response TDDFT-RPA\cite{AngelGWReview,response1,response2,DuncanGrapheneTDDFTRPA,Livia2014PSSB}, from the imaginary part of the macroscopic dielectric function, $\Imag[\varepsilon(\textbf{q},\omega)]$, in the limit $\|\textbf{q}\|\rightarrow 0^+$.  
Details of our implementation are provided in Appendix~\ref{LCAOTDDFTRPA}.

To model an excited singlet state of the system based on a fixed magnetic moment calculation, we ``swap'' between the spin channels ($s \in \{\uparrow,\downarrow\}$ or $\{0,1\}$) the eigenvalues and eigenfunctions of the levels beyond half the number of electrons, $N_e/2$. More precisely, we define
\begin{equation}
s' = \left\{
\begin{array}{ll}
s + 1\mod 2 &\text{if}\ n > N_{e}/2\\
s &\text{otherwise}
\end{array}\right.
\end{equation}
In this way we obtain the electronic structure of a singlet excited state that is constrained to have the same total electron density as the triplet ground state, $\rho_{\textit{ex}}^{S=0}(\textbf{r}) \equiv \rho_{\textit{gs}}^{S=1}(\textbf{r})$.  

Note that, as we are primarily interested herein with the absorption spectra near their onsets, local field effects may be neglected without impacting our results. This is demonstrated in Appendix~\ref{LCAOTDDFTRPA}, where we compare ground state TDDFT-RPA spectra for a (6,5) SWNT with and without including local field effects. 
\subsection{Computational Details}\label{ComputationalDetails}

All DFT calculations were performed with locally centered atomic orbitals (LCAOs) and the projector augmented wave (PAW) implementation within the \textsc{gpaw} code \cite{GPAW,GPAWRev,GPAWLCAO}. We used a double-zeta polarized (DZP) basis set for representing the density and the wave functions and the PBE exchange correlation (xc)-functional \cite{PBE}.  All calculations employed a room temperature Fermi filling ($k_B T \approx 25$~meV), with total energies extrapolated to $T\rightarrow 0$~K, i.e., excluding the electronic entropy contribution to the free energy $-ST$.  In this was we avoided an unrealistic smearing of the excited electron and hole in the triplet calculations.  We included \sfrac{2}{3}$N_e$ many bands within the calculations, which has been shown to be sufficient to converge the first $\pi\rightarrow\pi^*$ transitions within graphene \cite{DuncanGrapheneTDDFTRPA}.

Structural minimization was performed within the Atomic Simulation Environment (ASE) \cite{ASE}, until a maximum force below 0.05~eV/\AA\ was obtained. We employed more than 5~\AA\ of vacuum to the cell boundaries
 orthogonal to the (6,5) SWNT, (7,5) SWNT and polythiophene (PT),  and obtained optimized unit cells parameters of 40.92, 44.79, and 7.87~\AA\ along their axes, respectively.  Here, PT is modeled using two thiophene units in s-trans configuration.  We have used PT as a simplified model for P3HT, that is, removed the hexyl side chains of P3HT, since the influence of the P3HT side chains on the level alignment and charge transfer is negligible\cite{Livia2014PSSB,Imge}. Moreover, as we shall see in section~\ref{ResultsandDiscussion}, the influence of P3HT/PT on the IR spectrum of the SWNT is solely through hole-transfer.

\begin{figure}[!t]
\includegraphics[width=0.44\textwidth]{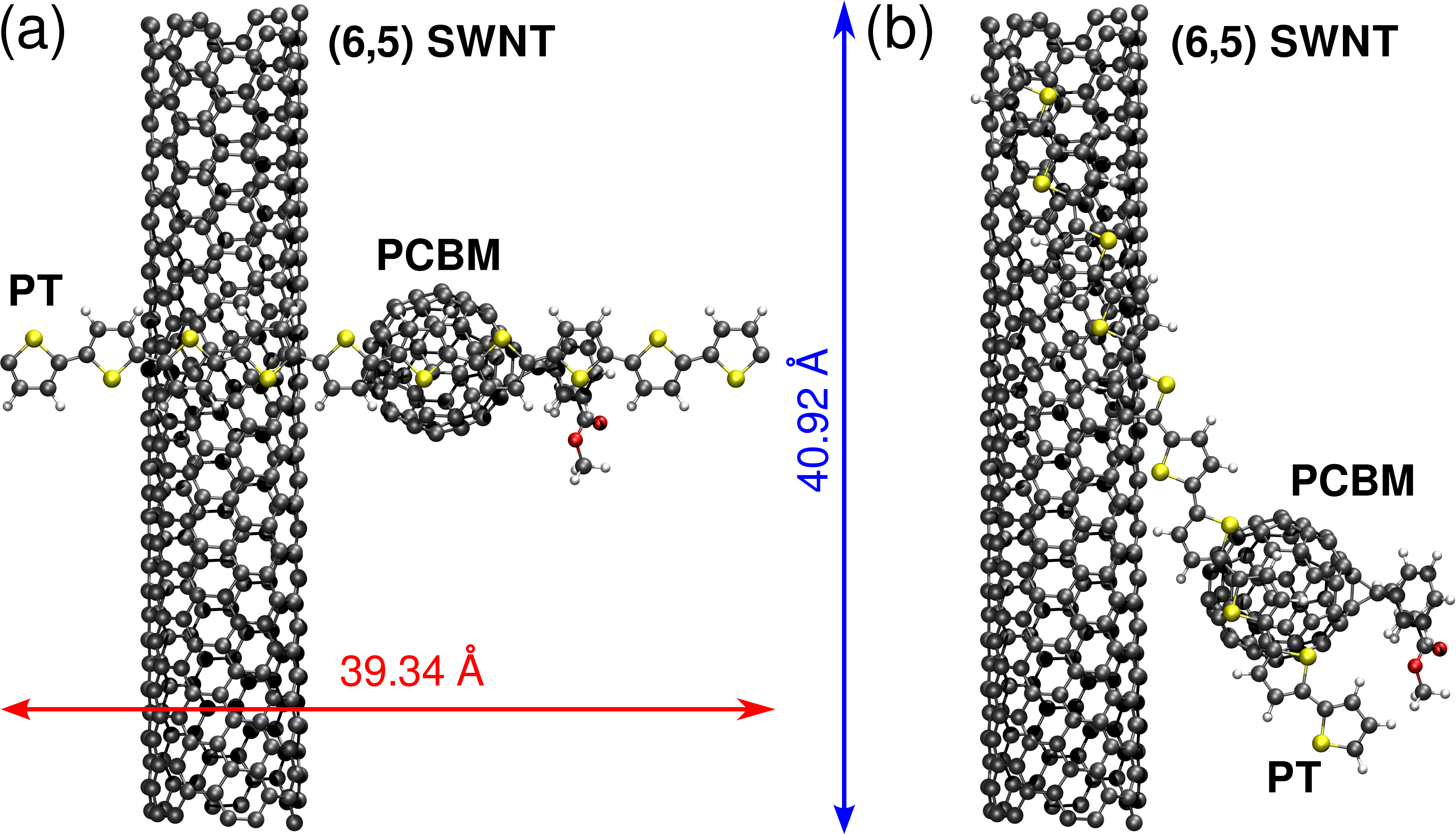}
\caption{Schematic of the PT/PCBM/(6,5) SWNT system with (a) a periodically repeated PT polymer aligned orthogonal to the SWNT and (b) a finite ten unit PT chain aligned with the SWNT. C, H, O, and S atoms are depicted by gray, white, red, and yellow balls, respectively.
}\label{fgr0}
\noindent{\color{StyleColor}{\rule{\columnwidth}{1pt}}}
\end{figure}

The PT/PCBM/(6,5)-SWNT bulk was modeled by a $39.34\times 40.92$~\AA$^2$ layered structure of ten thiophene units orthogonal to the SWNT axis, as shown in Figure~\ref{fgr0}(a).  In so doing, this configuration describes the limit of a minimal SWNT--PT overlap.  To determine the impact of increasing the SWNT--PT overlap, we also consider a truncated ten unit PT chain aligned with the SWNT and PCBM as shown in Figure~\ref{fgr0}(b).  

It has previously been shown that changes in the orientations of PCBM next to P3HT only cause energy differences within the accuracy of DFT\cite{PCBMorientations}. Since the potential energy surface is rather flat\cite{PCBMorientations}, we have chosen a smallest C--C intramolecular distance of $\sim3.3$~\AA\ between the relaxed P3HT, PCBM, and SWNT structures, and performed single-point calculations for the resulting configurations shown in Figure~\ref{fgr0}.  Although this C--C separation is 0.1~\AA\ smaller than the interlayer distance of multiwalled carbon nanotubes (MWNTs) and graphite \cite{graphite-layer}, it has been chosen to ensure an overlap between the SWNT and PT outer LCAO orbitals. An increase of the PT--SWNT/PCBM distance to $\sim3.4$~\AA\ changes the total energy by less than 50 meV, i.e., within the accuracy of DFT. In each case we found the repulsive forces for the close lying C atoms were  $\lesssim0.1$~eV/\AA.

\section{RESULTS AND DISCUSSION}\label{ResultsandDiscussion}

\subsection{Differential Transmission Measurements}\label{TransientAbsorptionMeasurements}

\begin{figure}[!t]
\includegraphics[scale=0.42]{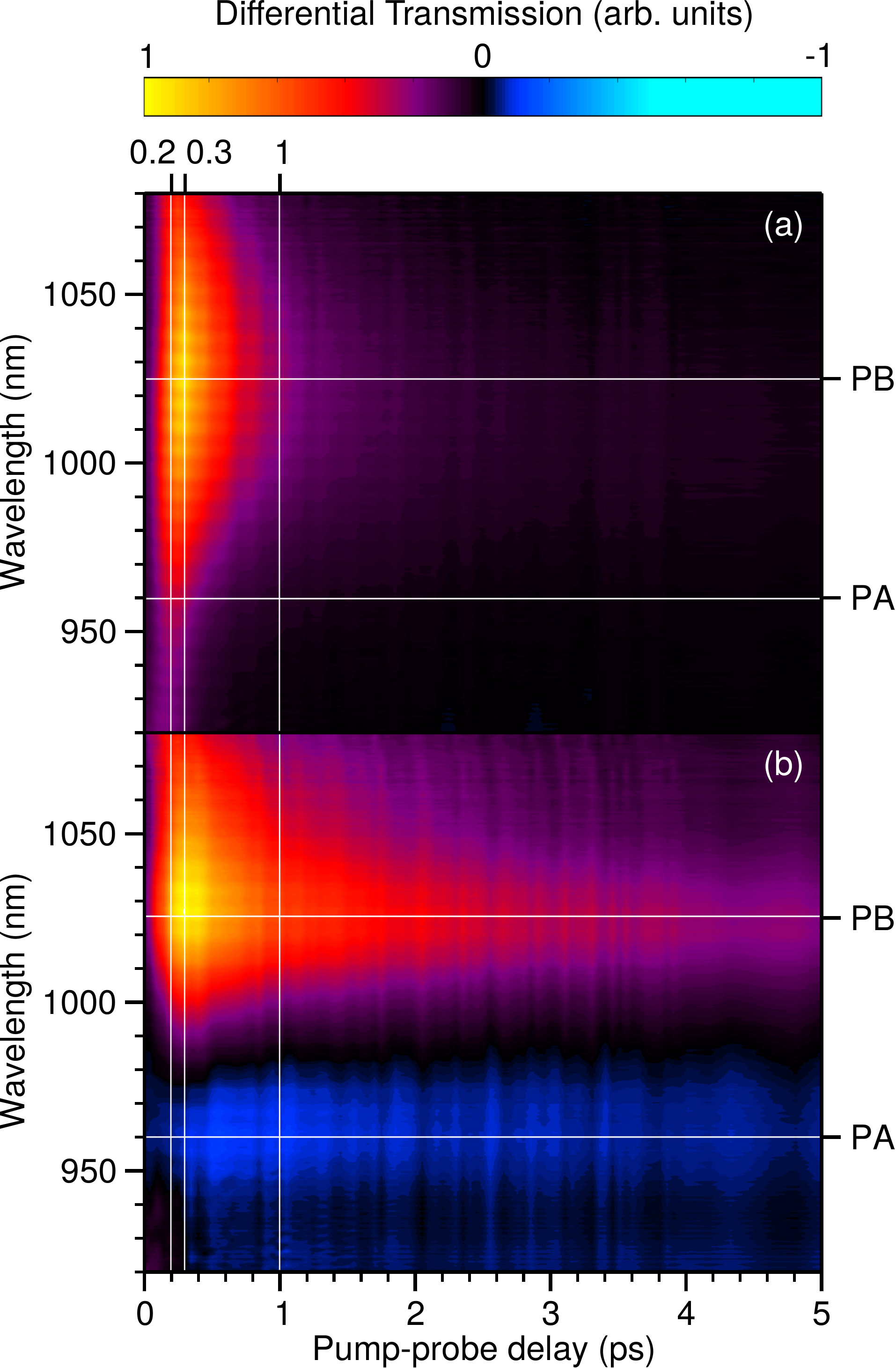}
\caption{Normalized differential transmission $\Delta T/T$ of (a) SWNT and (b) P3HT/PCBM/SWNT devices versus pump-probe delay in ps and probe wavelength in nm.  Pump-probe delays of 0.2, 0.3, and 1.0 ps (spectra in Figure~\ref{fgr2}) are marked by vertical lines, while photobleach (PB) and photoabsorption (PA) peaks (spectra in Figure~\ref{fgr3}) are marked by horizontal lines.
}\label{fgr1}
\noindent{\color{StyleColor}{\rule{\columnwidth}{1pt}}}
\end{figure}

\begin{figure}[!t]
\includegraphics[scale=0.42]{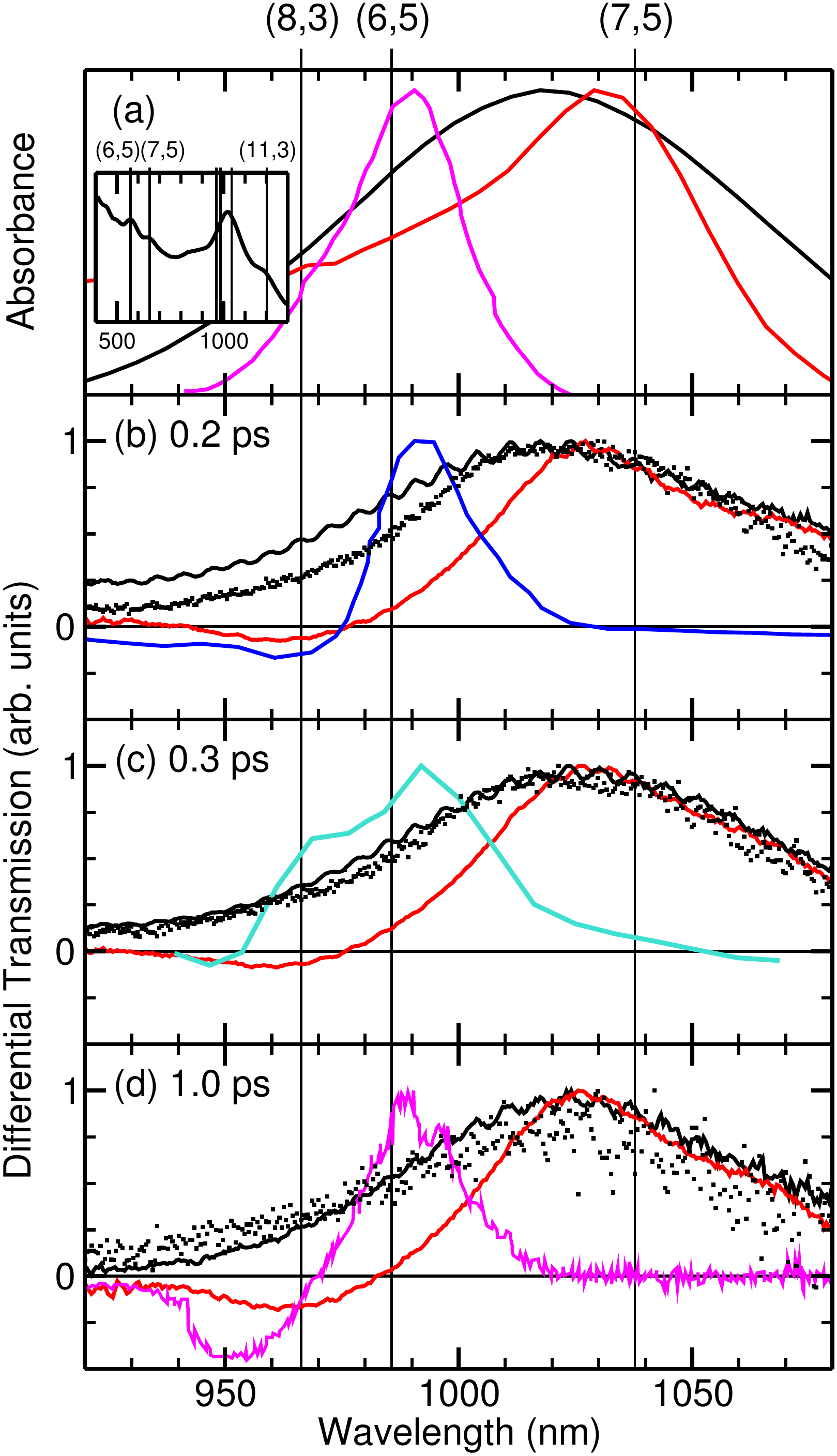}
\caption{Normalized (a) absorption and (b-d) differential transmission $\Delta T/T$ versus probe wavelength in nm for our SWNT sample (black) pumped at 900~nm (solid line) and 590~nm (squares), combined P3HT/PCBM/SWNT device (red) and from refs~\citenum{Zhuetal} (blue), \citenum{PRBKataura} (cyan), and \citenum{Zhu} (magenta) for pump-probe delays of (b) 0.2 ps, (c) 0.3 ps, and (d) 1.0 ps.  Wavelengths of the $E_{11}$ transitions for (8,3), (6,5), and (7,5) SWNTs from ref~\citenum{nature_nts} are marked above.
}\label{fgr2}
\end{figure}

Figure~\ref{fgr1}
shows the transient spectra of the SWNT and blended P3HT/PCBM/SWNT samples versus probing delay times up to 5 ps. In both spectra, we find a peak with a maximum around 1025~nm. Comparing the transient spectra with the absorbance (black) in Figure~\ref{fgr2}(a),
this peak can be assigned to the photobleach (PB) of mainly (6,5) and (7,5) SWNT $E_{11}$ transitions\cite{nature_nts}. For the blended sample, we have an additional photoinduced absorption (PA) peak around 960~nm. The samples show a remarkable difference in the decay rate of the PB peak. In the case of the SWNT sample, the $E_{11}$ transitions seem to be completely accessible again after 1 ps, while the decay of the $E_{11}$ exciton in the blended sample is 4--5 times longer.

The UV/VIS/NIR absorption spectra of the SWNT sample are shown as an inset to Figure~\ref{fgr2}(a). The $E_{22}$ transitions of (6,5) and (7,5) SWNTs are clearly visible, with the $E_{11}$ transition of (11,5) SWNTs  seen as a shoulder.

By comparing the transient spectra of the SWNT sample (black) at delay times of 0.2, 0.3, and 1 ps with the spectra of similar SWNT samples of the literature at the same delay times, we found a  discrepancy within the measured peak width and structure of the SWNT sample spectrum and those of the literature\cite{Zhuetal,Zhu,PRBKataura}, as shown in Figure~\ref{fgr2}(b-d). The (6,5) SWNT enriched samples of refs~\citenum{Zhuetal} (blue) and \citenum{Zhu} (magenta) show a much smaller peak width and an additional PA peak between 940 and 975 nm. The PA peak in the sample of ref \citenum{PRBKataura} (cyan) is overlaid by the PB of the (8,3) SWNT ($\sim966$~nm)\cite{nature_nts} and therefore not strongly pronounced. As well, the width is broader due to the mixture of tubes. This is consistent with the differential transmission of our SWNT sample.

To determine whether the pump energy has an influence on the peak width, the SWNT sample was pumped at two different wavelengths: 590 nm in the VIS and 900 nm in the IR. These pump energies are comparable to those reported the literature\cite{PRBKataura,Zhuetal,Zhu}.  For example, ref~\citenum{PRBKataura} pumped in the infrared region at 930 nm, while the pump beams at 570 nm employed in refs~\citenum{Zhuetal} (blue) and \citenum{Zhu} (magenta) were tuned to the $E_{22}$ transition.

As shown in Figure~\ref{fgr2}(b,c,d), the differential transmission spectra for our SWNT sample is rather insensitive to whether pumping is in the VIS or IR.  This suggests that the broader PB peak within our SWNT sample, shown in Figure~\ref{fgr2}(a), is related to the sample itself.    It has previously been argued that SWNT mixtures have an increased exciton transfer and electron-hole trapping between tubes\cite{Holt,Strano}. Thus, the greater PB peak width in our SWNT sample may be related to having a mixture of both (6,5) and (7,5) SWNTs.  However, having a SWNT mixture does not explain the occurrence of the PA peak in the blended P3HT/PCBM/SWNT sample (red). 

Taking a closer look at the transient spectrum of the SWNT sample in Figure~\ref{fgr2}(b,c,d), we notice it exhibits a small asymmetry. Nevertheless, including a range of $\pm$50 nm from the peak maximum, this asymmetry cannot be completely verified, as it is within the noise of the measurement. Thus, the PA appears to completely vanish in the spectra of the SWNT sample.

\begin{figure}[!t]
\includegraphics[scale=0.42]{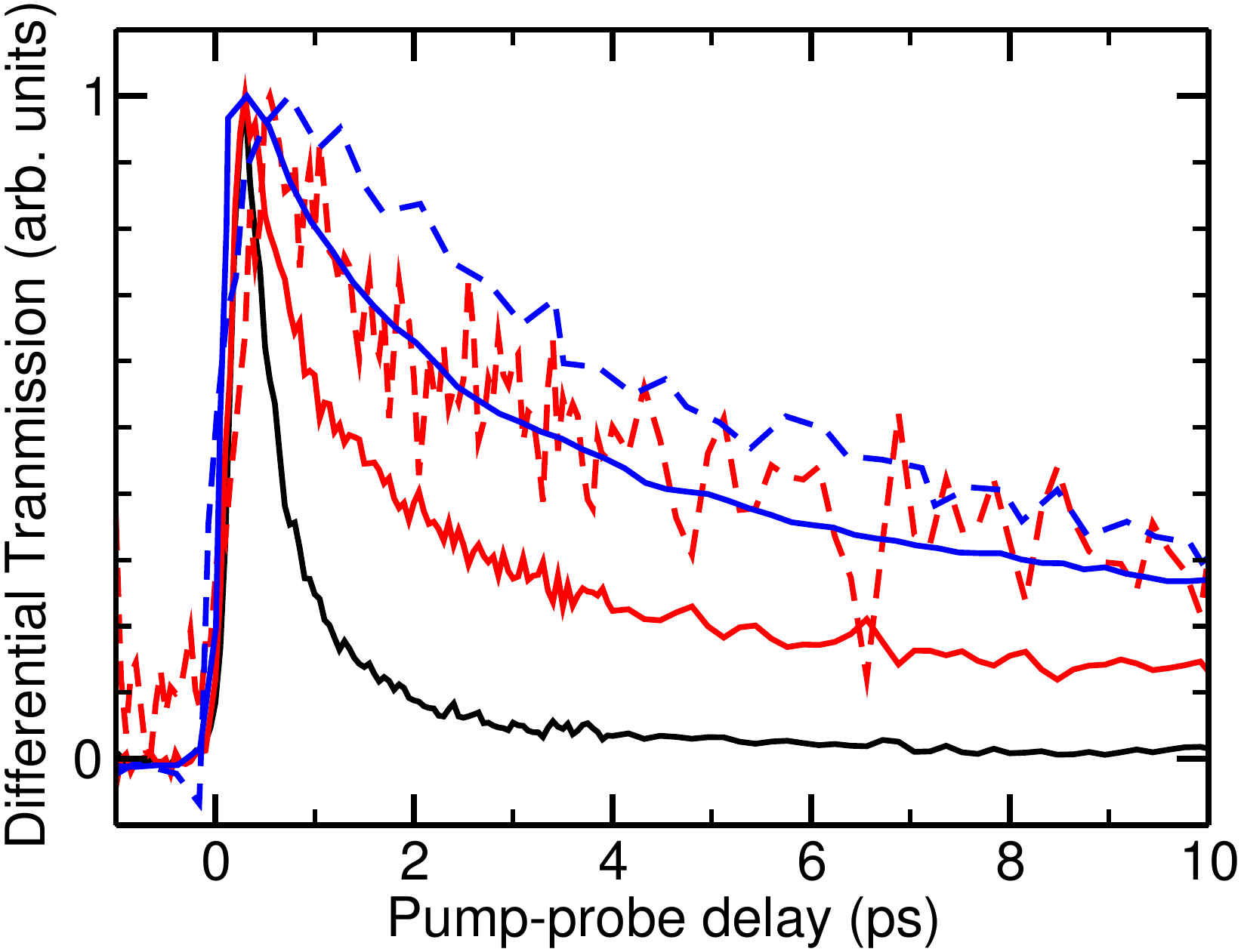}
\caption{Normalized differential transmission $\Delta T/T$ as a function of the pump-probe delay in ps measured at the maximum of the photobleach (PB) peak (solid lines) and photoabsorption (PA) peak (dashed lines) for our SWNT sample (black), combined P3HT/PCBM/SWNT device (red), and for (6,5) SWNTs from ref~\citenum{Zhuetal}.}\label{fgr3}
\noindent{\color{StyleColor}{\rule{\columnwidth}{1pt}}}
\end{figure}

The recovery time dynamics of the PB and PA peaks for the mixed SWNT sample and the P3HT/PCBM/SWNT blended sample are shown in Figure~\ref{fgr3}.
We find the mixed SWNT sample's recovery time is much shorter than that of the blended P3HT/PCBM/SWNT sample, with the recovery time reported in ref~\citenum{Zhuetal} for a predominantly (6,5) SWNT sample even longer.  In each case, the PA peaks seem to recover as fast as the PB peaks or slightly slower.

The main reason for the extinction of the PA peak is the large peak width, which is crucial for the visibility of the PA peak. However this does not explain its origin. 

\subsection{Interpreting differential transmission as a Difference Spectra}\label{InterpretingTransientAbsorptionasaDifferenceSpectra}

Zhu et al.\cite{Zhuetal} found a PA peak, and suggested it is due to a biexcitation. In a subsequent paper, Zhu \cite{Zhu} suggested coupling to the radial breathing mode as a reason for the PA peak.

\begin{figure}
\noindent{\color{StyleColor}{\rule{\columnwidth}{1pt}}}
\includegraphics[scale=0.42]{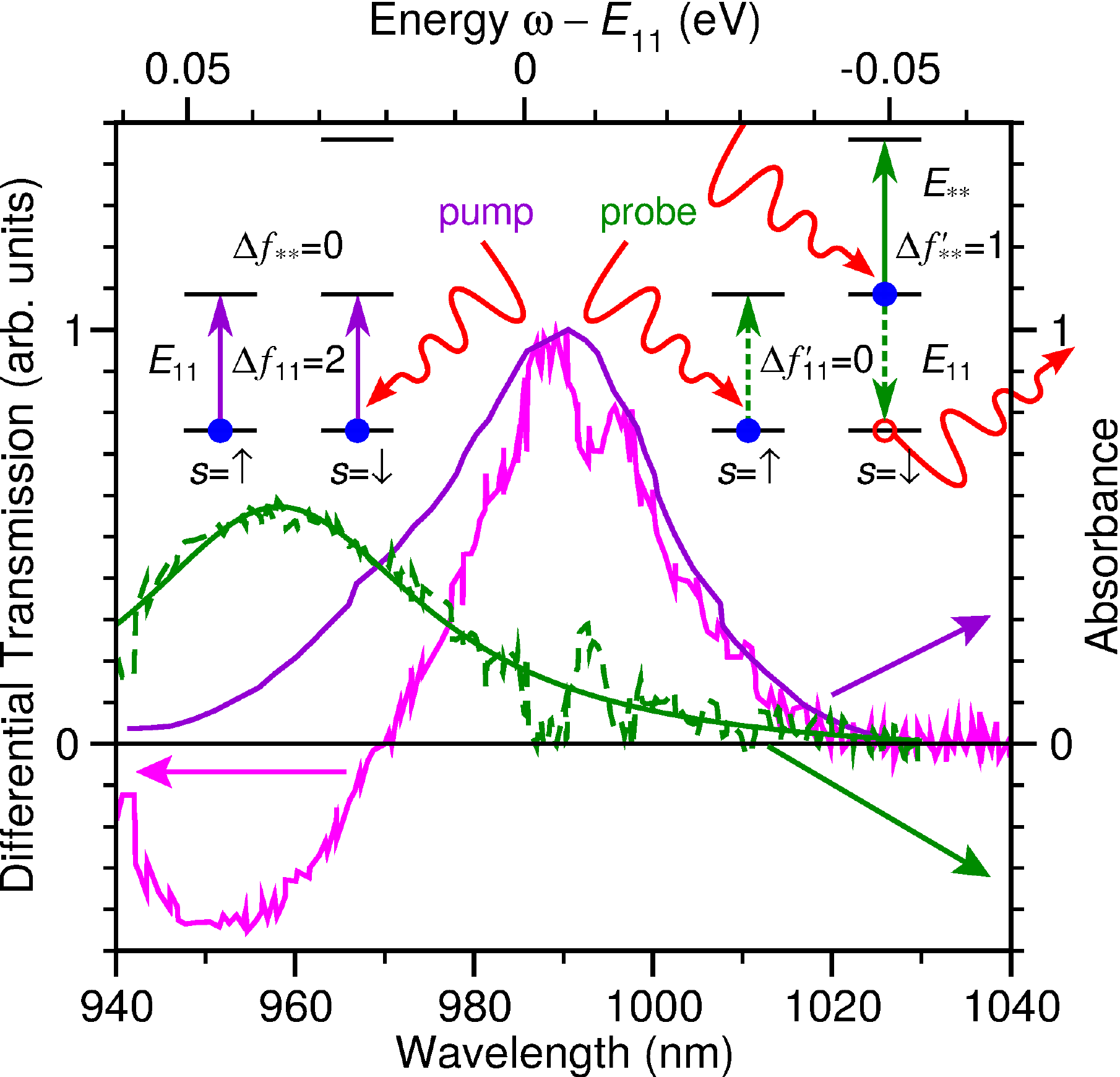}
\caption{Normalized initial absorbance (violet) and differential transmission $\Delta T/T$ with a pump-probe delay of 1 ps (magenta) from ref~\citenum{Zhu}, and their difference (green) versus probe wavelength in nm and energy in eV relative to the $E_{11}$ transition. A Lorentzian fit to the difference between the initial absorbance and differential transmission is also shown. An excited electron transition $E_{**}$ model for the measured absorbance/differential transmission is shown schematically in the upper left/right inset.
}\label{fgr4}
\noindent{\color{StyleColor}{\rule{\columnwidth}{1pt}}}
\end{figure}

Evaluating the experimental data of ref \citenum{Zhu} shown in Figure~\ref{fgr4},
we identified the peak causing the PA peak by taking the difference between the normalized absorbance (violet) and the normalized transient spectrum (magenta). This results in the difference spectrum of the probe beam (green), when the system is fully excited by the pump beam, that is, all electrons of the $E_{11}$ transition are in the CBM. In this case, the excited system gives rise to an absorbance of more than 50\% of the PB peak but at the energy of the PA peak.

Suppose the PA peak is due to absorption by an excited electron, i.e., an excited electron transition $E_{**}$, as depicted schematically in Figure~\ref{fgr4}. It should be noted that the single particle picture employed here refers to the system being excited, rather than the excitation process itself.  When the system is initially excited by the pump, the difference in filling for the $E_{11}$ transition, $\Delta f_{11} = 2$, while the difference in filling for the $E_{**}$ transition's levels $\Delta f_{**} = 0$. When the system is subsequently excited by the probe, that is, in the presence of an exciton, $\Delta f_{11}' = 0$, while $\Delta f_{**}' = 1$.  Since a transition's intensity is proportional to the difference in filling, the 2:1 ratio between PB and PA peaks is already accounted for by $\Delta f_{11}:\Delta f_{**}'$.  This requires the $E_{11}$ and $E_{**}$ transitions to have almost the same overlaps.  

Although this does not rule out the possibility of an $E_{**}$ transition being responsible for the PA peaks close to the PB peaks within the SWNT systems, it makes it rather unlikely.  More importantly, it suggests the PA and PB peaks most likely arise from the same $E_{11}$ transition.  In fact, in section~\ref{TDDFTSimulatedSpectra} we will show that the $E_{11}$ transition is blue shifted by a band gap widening in the excited state, potentially explaining the origin of the PA peak. 

\subsection{Modeling Optically Excited States}\label{ModelingOpticallyExcitedStates}

As discussed in section~\ref{TheoreticalMethods}, we model the optically excited state using a singlet excited state whose total electron density is constrained to be that of the triplet ground state.  This singlet excited state is a suitable approximation to the optically excited state if the spin densities associated with the Kohn Sham (KS) valence band maximum (VBM) and conduction band minimum (CBM) eigenfunctions, $|\psi_{\textrm{VBM} s}|^2$ and $|\psi_{\textrm{CBM} s}|^2$, are only weakly dependent on the spin channel $s$.  In other words, if the two spin channels in the triplet DFT calculation are basically equivalent up to a phase factor, i.e., $|\langle\psi_{n\uparrow}^{S=1}|\psi_{n'\downarrow}^{S=1}\rangle| \approx \delta_{n n'}$, the optically excited state should have a similar electron density.  If, moreover, the singlet and triplet ground state eigenfunctions are also basically equivalent up to a phase factor, i.e., $|\langle\psi_{n\textit{gs}}^{S=1}|\psi_{n'\textit{gs}}^{S=0}\rangle| \approx \delta_{n n'}$, this singlet excited state should describe the optically excited state quite well.

We find this is indeed the case for the (6,5) and (7,5) SWNTs, with the KS eigenfunctions having approximately the same spatial distribution in the singlet and triplet ground states.  For the blended SWNT/PT/PCBM systems, the KS eigenfunctions have similar spatial distribution in both spin channels for the triplet DFT calculation, but differ from the singlet ground state KS eigenfunctions.  In particular, the occupied CBM level of the triplet DFT calculation is a hybridization of the first three CB eigenfunctions from the singlet ground state calculation.  This is indicative of charge transfer in the blended SWNT/PT/PCBM system.

Overall, these results strongly suggest that the total electron density of the triplet DFT calculation should be quite similar to that of the optically excited state.  This justifies our use of the singlet excited state electronic structure, obtained by "swapping" between spin channels the CB eigenvalues and eigenfunctions of the triplet ground state, to model the optically excited state.

\subsection{TDDFT-RPA Simulated Spectra}\label{TDDFTSimulatedSpectra}

\begin{figure}[!h]
\noindent{\color{StyleColor}{\rule{\columnwidth}{1pt}}}
\includegraphics[scale=0.42]{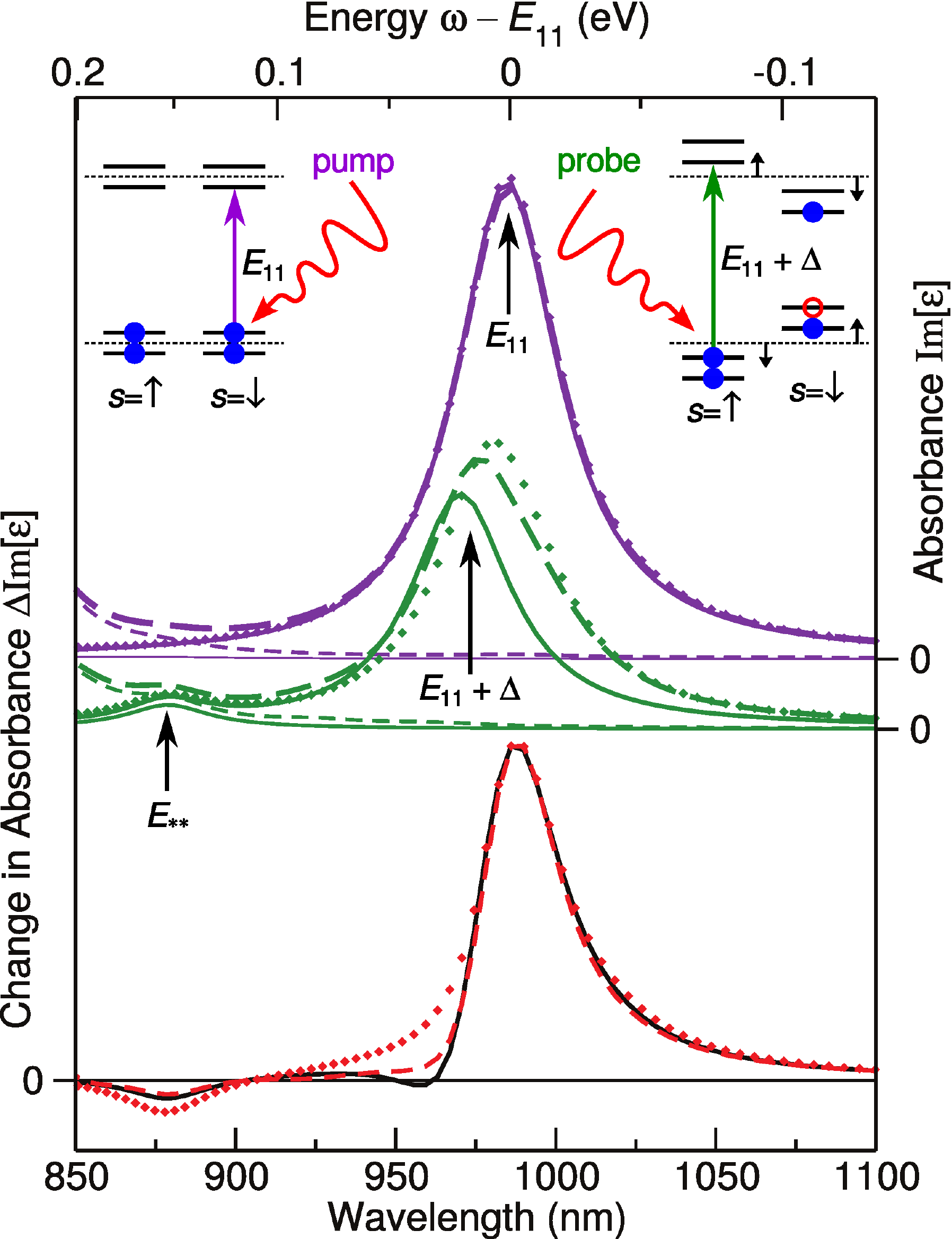}
\caption{The TDDFT-RPA absorbance $\Imag[\varepsilon]$ of a (6,5) SWNT (solid lines) and for the combined PT/PCBM/(6,5)~SWNT system depicted in Figure~\ref{fgr0}(a) (dashed lines) and (b) (diamonds) in the ground/excited state by the pump/probe (violet/green) as depicted schematically in the upper left/right inset, and  change in absorbance between the ground and excited state for a (6,5) SWNT (black) and a combined PT/PCBM/(6,5) SWNT system (red) versus wavelength in nm and energy in eV relative to the $E_{11}$ transition. Total absorbance (thick lines) and absorbance for light polarized perpendicular to the SWNT axis (thin lines) are shown.}\label{fgr5}
\noindent{\color{StyleColor}{\rule{\columnwidth}{1pt}}}
\end{figure}

The differential transmissions in the long term limit of the (6,5) SWNT (black) and  bulk PT/PCBM/(6,5) SWNT systems (red) in Figure~\ref{fgr5}
are calculated using the difference between the TDDFT-RPA ground state (violet) and excited state (green) absorbances. The ground state absorbance of the bulk system (dashed) shows the onset of PT absorbance at $\sim925$~nm, orthogonal to the $E_{11}$ transition.  In the range of 925 to 1100~nm, both systems, the (6,5) SWNT and the bulk, overlap. This suggests that neither PT nor PCBM transitions are involved within this energy range, and there are only (6,5) SWNT transitions. The excited state absorbances (green) of both systems show two significant changes: the PB peak is shifted to higher energy and there is a new transition at $\sim880$~nm. This new peak with rather low intensity is absorbing orthogonal to the tube axis, and can be assigned to an intratube interband transition $E_{**}$ of the excited electron in the CBM to an energy level $\sim1.1$~eV above the CBM.  However, this peak vanishes when local field effects are included, as shown in Appendix~\ref{LCAOTDDFTRPA}, which may be related to modeling debundled SWNTs \cite{SWCNTbundleRPA}.

The schematic in Figure~\ref{fgr5} explains the origin of the $E_{11}$ transition shifts in the excited states. As discussed in section~\ref{TheoreticalMethods}, we modeled the excited state by computing a triplet state in the first step. The singlet excited state is obtained by swapping the eigenenergies and eigenfunctions between spin channels for the originally unoccupied states, i.e., $n >N_e/2$. In this way, the electron and the hole are arranged in the same spin channel. As a result, the bands including the electron and hole are shifted closer in energy (electron-hole binding), while the other channel with the electron in
the valence band (VB) increases its band gap. This is not seen in the triplet calculation itself. Due to this widening of the band gap, the second excitation of an electron in the VB is at higher energy.

In the changes of the absorbance, i.e., differential transmission in the long term limit, the $E_{**}$ transitions give rise to negative peaks. However, they are at much higher energy than the PA peaks measured experimentally. On the other hand, the shifts $\Delta$ of the $E_{11}$ transitions in the excited state cause a reduction of the transient absorbance peaks at $\sim960$~nm. The larger shift within the (6,5) SWNT system even leads to a small negative peak in the change in absorption. Overall, the (6,5) SWNT difference spectrum agrees qualitatively with that of ref~\citenum{Zhuetal}, shown in Figure~\ref{fgr2}(b).

This is exactly the opposite from what we see in the experimental data obtained from the SWNT and the blended samples. There, we have no PA in the transient spectrum of the SWNT sample, but a more pronounced reduction of the peak at the higher energy end for the blended sample spectrum.  To understand the origin of this discrepancy, we consider the role of charge carrier loading, i.e., addition and removal of electrons and holes, and peak broadening on the differential transmission spectra in the following sections.

\subsection{Role of Charge Carrier Loading}\label{InfluenceofCharging}

\begin{figure}[!h]
\noindent{\color{StyleColor}{\rule{\columnwidth}{1pt}}}
\includegraphics[width=\columnwidth]{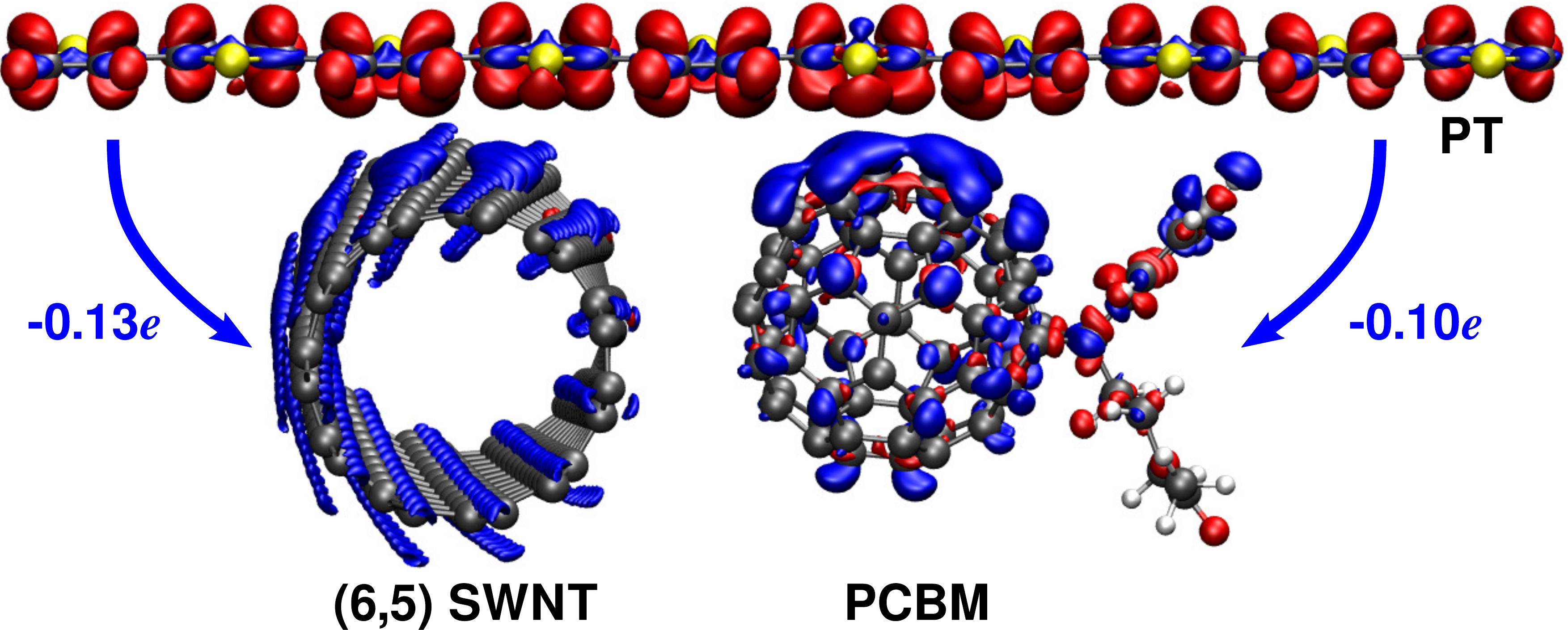}
\caption{Electron (blue) and hole (red) densities for a combined PT/PCBM/(6,5) SWNT system from the DFT total electron density difference between the excited and ground state. The charge transfer upon excitation of $-0.13e$ from the PT to the (6,5) SWNT and of $-0.10e$ from PT to PCBM is depicted schematically. C, H, O, and S atoms are depicted by gray, white, red, and yellow balls, respectively.}\label{fgr7}
\noindent{\color{StyleColor}{\rule{\columnwidth}{1pt}}}
\end{figure}

We calculated the excited electron (blue) and hole (red) densities as the difference between the electron density in the ground state and triplet configuration, as shown in Figure~\ref{fgr7}.

\begin{figure*}[!b]
\noindent{\color{StyleColor}{\rule{\textwidth}{1pt}}}
\includegraphics[width=\textwidth]{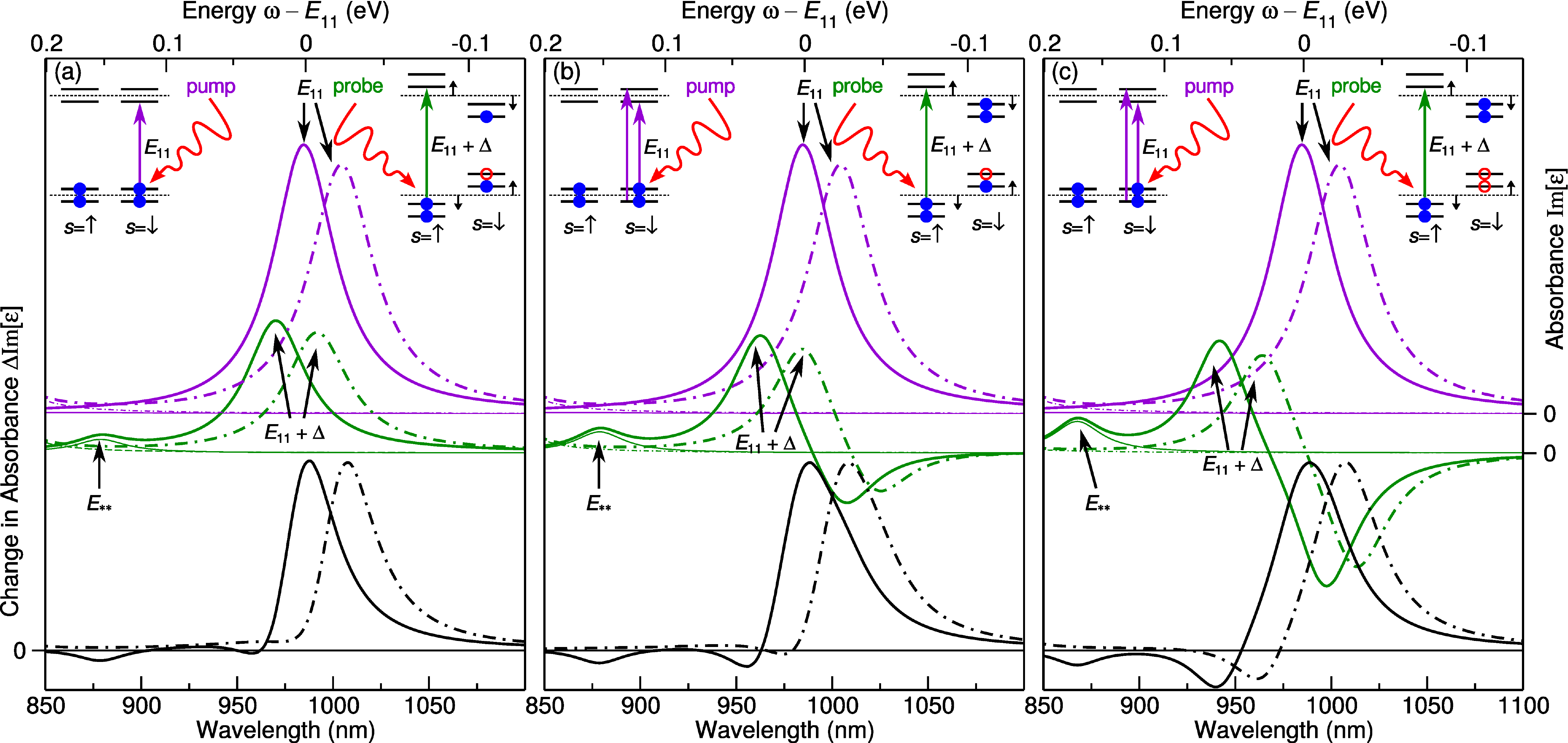}
\caption{TDDFT-RPA absorbance $\Imag[\varepsilon]$ of a (6,5) SWNT (solid lines) and (7,5) SWNT (dashed-dotted lines) in the ground/excited state by the pump/probe (violet/green) as depicted schematically in the upper left/right inset, and change in absorbance between the ground and excited state versus wavelength in nm and energy in eV relative to the $E_{11}$ transition of a (6,5) SWNT. The excited state is modeled by (a) a single exciton, (b) a negative trion, and (c) a pair of excitons. Total absorbance (thick lines) and absorbance for light polarized perpendicular to the SWNT axis (thin lines) are shown.
}\label{fgr8}
\end{figure*}

In this way we include all electron redistributions in the VB and CB.  The difference in the densities reveals a hole located mainly on P3HT/PT, whereas the electron is predominantly on the (6,5) SWNT and the PCBM. This results in a net charge transfer of 0.23 electrons from the PT to its neighboring molecules. The amount of calculated charge transfer within the bulk system is in agreement with previous DFT results for P3HT/fullerene and P3HT/SWNT heterojunctions\cite{DFT-P3HT-Fullerene, DFT-P3HT-Tubes}.

The charge transfer within the bulk system partially fills the hole in the excited state. This transfer of charge from PT into the (6,5) SWNT VBM stabilizes it, making the band gap widen less than in the isolated (6,5) SWNT system. Even though the transition is more intense, due to the additional electrons, the reduction of the PB peak at the high energy end is less pronounced due to the smaller shift $\Delta$.

To determine how the amount of hole transfer from the SWNT and higher loading of charge carriers influences the  shape of the transient spectrum at the $E_{11}$ transition of SWNT systems, we have increased the amount of charge carriers on the (6,5) and (7,5) SWNTs from 0.24 and 0.22 to 0.49 and 0.45 $e$/nm, respectively. In Figure~\ref{fgr8} we compare the change in TDDFT-RPA absorption $\Imag[\varepsilon]$ (from eq \ref{Imeps} of Appendix~\ref{LCAOTDDFTRPA}) of (6,5) and (7,5) SWNTs  with (a) a single exciton, (b) a negative trion, i.e., a pair of excited electrons and a single hole, and (c) a pair of excitons.  Unsurprisingly, with the addition of a second exciton, as shown in Figure \ref{fgr8}(c), the widening of the band gap $\Delta$ is significantly increased, resulting in a stronger PA peak at higher energy ($\sim940$~nm for the (6,5) SWNT). These results are consistent with the predominantly (6,5) SWNT measurements of ref~\citenum{Zhu}  after a 1 ps pump-probe delay, as shown in Figure~\ref{fgr2}(d). There, the PA peak is at $\sim950$~nm and even gains half of the intensity of the PB peak. Additionally, the existence of a second exciton within a 4~nm unit cell is quite reasonable, as compared to the calculated exciton size of $\sim2$~nm \cite{larry-exlength}. In Figure \ref{fgr8}(b) one of the holes is filled, e.g., through charge transfer from P3HT to a SWNT, the PB peak becomes more asymmetric, and the PA peak becomes less intense. These results are more in agreement with the measurements after a 0.2 ps pump-probe delay of ref~\citenum{Zhuetal} shown in Figure~\ref{fgr2}(b). Overall, this suggests the strength of the PA peak, and the degree to which it is blue shifted from the PB peak $\Delta$, may be used as qualitative measures of the charge transfer and charge carrier load within SWNT systems.

Unfortunately, this still does not explain, why the blended P3HT/PCBM/SWNT sample has a more pronounced PA peak, since actually we would expect the opposite considering the amount of charge carriers, that is, the amount of hole, on the SWNT.

Comparing the calculated change in absorbance for (6,5) and (7,5) SWNTs in Figure~\ref{fgr8}, we find the (7,5) SWNT exhibits a less pronounced PA peak than the (6,5) SWNT for a single exciton, a negative trion, and a pair of excitons.  In each case, the (7,5) SWNT excited state calculation gave a smaller gap widening $\Delta$ than the (6,5) SWNT.  More importantly, the (7,5) SWNT PA peak overlaps with the PB peak of the (6,5) SWNT.  This suggests that for a mixture of (6,5) and (7,5) SWNTs, a PA peak will only be visible when excited electrons remain on the (6,5) SWNT.

\subsection{Influence of PB Peak Width}\label{InfluenceofPeakWidth}

It was already shown that the intensity and visibility of a PA peak decreases, if the broadening is too high \cite{width-theory}. To see what happens with the width and the shape of the peak for different broadenings, we modeled the SWNT sample spectrum by combining the calculated (6,5) and (7,5) SWNT transient spectra for various broadenings $\Gamma$.

\begin{figure}[!t]
\includegraphics[scale=0.42]{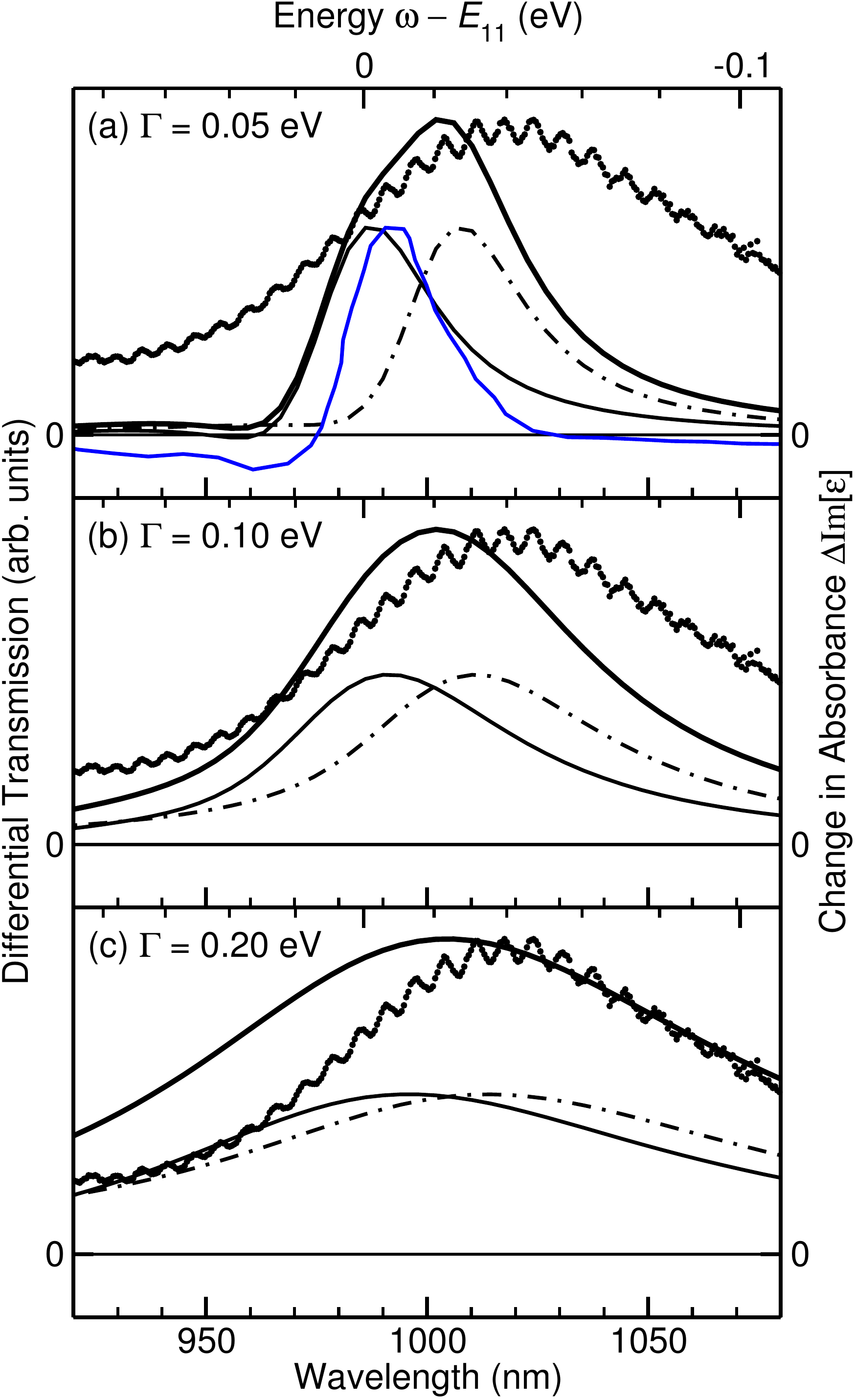}
\caption{TDDFT-RPA change in absorbance between the ground and excited state for a (6,5) SWNT (solid thin lines), (7,5) SWNT (dash-dotted lines), and their normalized sum (solid thick lines) for broadenings $\Gamma$ of (a) 50 meV, (b) 100 meV, and (c) 200 meV versus wavelength in nm and energy in eV relative to the $E_{11}$ transition.  Normalized differential transmission $\Delta T/T$ for our SWNT sample (black dots) and from ref~\citenum{Zhuetal} (blue) for a pump-probe delay of 0.2 ps also provided.}\label{fgr6}
\noindent{\color{StyleColor}{\rule{\columnwidth}{1pt}}}
\end{figure}
In Figure~\ref{fgr6}  the combined (6,5) and (7,5) SWNT differential transmission spectra for peak widths of 0.05, 0.1, and 0.2 eV for the ground state and excited state absorbance are shown. A peak width of 0.05 eV for the independent tube absorbances is in good agreement with the peak width of the transient spectrum of ref~\citenum{Zhuetal} (blue). A peak width of 0.05 eV is also consistent with the absorption FWHM of ref~\citenum{Zhu}, shown in Figures~\ref{fgr2}(a). However, in comparison to the spectrum of our SWNT sample, the combined transient spectrum has a much smaller peak width and is shifted to higher energy. This suggests that the SWNT sample shows a significant contribution due to (7,5)  SWNTs.  Furthermore, the transient spectrum of ref~\citenum{Zhuetal} shows a clear asymmetry due to the reduction of the peak at the higher energy end, consistent with our TDDFT-RPA calculations.  This asymmetry becomes less pronounced the greater the broadening of the transitions. A peak width of 0.2~eV results in a spectrum similar to our SWNT sample measurements. This is also consistent with the absorption FWHM of our SWNT mixture shown in Figure~\ref{fgr2}(a).

A broader peak width for the SWNT and blended P3HT/PCBM/SWNT samples compared to those in the literature\cite{Zhu,Zhuetal,PRBKataura} might be due to haveing (6,5) and (7,5) SWNT mixture. Exciton transfer from the (6,5) to the (7,5) SWNT may be expected to occur, leading to electron-hole trapping on the (7,5) SWNT. Already, a 20\% impurification can cause a decrease in power conversion efficiency of an OPV by more than 30 times due to electron-hole trapping \cite{Strano}.  However, the peak width of the blended sample is significantly narrower than that of the SWNT system, as seen in Figure~\ref{fgr2}(a). The calculated energy differences between the excited (triplet) and ground states of the (6,5) SWNT and the bulk PT/PCBM/SWNT system suggest that an exciton within the bulk is more stable by about ~0.08 eV. A potential source for the increased stability, and hence exciton lifetime, for the bulk system is electron and hole delocalization. This might explain the smaller width of the blended P3HT/PCBM/SWNT absorbance and resulting increase in visibilty of the PA peak.

\section{CONCLUSIONS}\label{Sect:Conclusions}

Using linear response TDDFT-RPA calculations of the ground and excited states of (6,5) and (7,5) SWNTs, we are able to qualitatively explain the measured differential transmission spectra of the probed lowest energy excited state, the $E_{11}$ transition. Our results confirm that the observed PA peak is due to a blue shift $\Delta$ of the $E_{11}$ transition after pumping. The PA peak is therefore an artifact of excitations within the spin channel, which experience a widening of the band gap after electron and hole stabilization within the other spin channel.

The intensity and the visibility of the PA peak depends on the charge carrier density, but mainly on the peak width. If the peak width is greater than the blue shift of the main absorption peak, the transient spectrum will be dominated by the PB peak\cite{width-theory}. 
Further, our SWNT sample contains a mixture of (6,5) and (7,5) SWNTs. This SWNT sample exhibits rather broad absorption peaks, approximately 4 times broader than previously measured (6,5) SWNT samples \cite{Zhuetal,Zhu}. We suggest that this is related to having a mixture of SWNTs with similar energy gaps. The different types of SWNTs are causing an overlap of the peaks, but more importantly, the mixture enables an exciton transfer from the (6,5) to (7,5) SWNTs. This both shortens the exciton lifetime on the (6,5) SWNTs and inhibits a widening of the band gap. As a result, the differential transmission spectrum of the (6,5) SWNT has almost no PA peak.

For the blended P3HT/PCBM/SWNT sample, the exciton is locally stabilized on the (6,5) SWNT due to the hole transfer to the P3HT. As a result, the PA peak intensity and position in the differential transmission spectra may be used as a qualitative measure of exciton density and charge transfer within SWNT systems.

These results are important for the understanding of the origin of the PA peak in pump-probe spectroscopy and will help to interpret the exciton dynamics within SWNT systems. 

\appendix

\section{LCAO TDDFT-RPA}\label{LCAOTDDFTRPA}

The optical absorption spectra are obtained via linear response time dependent (TD) density functional theory (DFT) within the random phase approximation (RPA)\cite{AngelGWReview,response1,response2,DuncanGrapheneTDDFTRPA,Livia2014PSSB}, from the imaginary part of the macroscopic dielectric function, $\Imag[\varepsilon(\textbf{q},\omega)]$, as $\|\textbf{q}\|\rightarrow 0^+$.  
In general, within linear response TDDFT-RPA\cite{AngelGWReview,response1,response2}, the dielectric matrix in reciprocal space is given by
\begin{equation}
\varepsilon_{\textbf{G}\textbf{G}'}(\textbf{q},\omega) = \delta_{\textbf{G}\textbf{G}'} - \frac{4\pi}{\|\textbf{q}+\textbf{G}\|^2}\chi_{\textbf{G}\textbf{G}'}^0(\textbf{q},\omega),\label{epsilonmatrix}
\end{equation}
where $\textbf{G}$ and $\textbf{G}'$ are reciprocal lattice vectors, and $\chi_{\textbf{G}{\textbf{G}'}}^0$ is the noninteracting density--density response function, i.e., the susceptibility.  This is given by
\begin{eqnarray}
\chi_{\textbf{G}\textbf{G}'}(\textbf{q},\omega)\!\!\!&=&\frac{1}{\Omega}\sum_{\textbf{k}}\sum_{n m}\sum_{ss'}\frac{f_{ns'\textbf{k}} - f_{ms\textbf{k}+\textbf{q}}}{\omega + \varepsilon_{ns'\textbf{k}} - \varepsilon_{ms\textbf{k+q}} + i\eta}\label{chi0}\\
&\times&\langle\psi_{ns'\textbf{k}}|e^{-i(\textbf{q}+\textbf{G})\cdot\textbf{r}}|\psi_{ms\textbf{k}+\textbf{q}}\rangle\langle\psi_{ns'\textbf{k}}|e^{i(\textbf{q}+\textbf{G}')\cdot\textbf{r}'}|\psi_{ms\textbf{k}+\textbf{q}}\rangle\nonumber
\end{eqnarray}
where $\eta$ is the electronic broadening, i.e., twice the inverse lifetime $\Gamma$ of the transitions, $\Omega$ is the supercell volume, $f_{ns'\textbf{k}}$ is the Fermi-Dirac occupation, $\varepsilon_{ns'\textbf{k}}$ is the eigenenergy, and $\psi_{ns'\textbf{k}}$ is the KS wave function of the $n$th band in spin channel $s'$ at $k$-point \textbf{k}.

For $\textbf{G} = \textbf{G}' = 0$, in the limit $\|\textbf{q}\|\rightarrow 0^+$, the matrix elements in (eq~\ref{chi0}) reduce to
\begin{equation}
\langle\psi_{ns'\textbf{k}}|e^{-i(\textbf{q}+\textbf{G})\cdot\textbf{r}}|\psi_{ms\textbf{k}+\textbf{q}}\rangle = -i \textbf{q}\cdot \frac{\langle\psi_{ns'\textbf{k}}|\nabla|\psi_{ms\textbf{k}+\textbf{q}}\rangle}{\varepsilon_{ns'\textbf{k}} - \varepsilon_{ms\textbf{k}}}.\label{matrixelements}
\end{equation}

Including local field effects, the absorption is obtained from (eq~\ref{epsilonmatrix}) by solving the Dyson equation
\begin{equation}
\Imag[\varepsilon(\omega)] = \lim_{\|\textbf{q}\|\rightarrow0^+}\frac{1}{\varepsilon^{-1}_{00}(\textbf{q},\omega)}.
\end{equation}

Neglecting local field effects, parallel to $\textbf{q}$, i.e., $\hat{e}_\textbf{q}$, $\varepsilon(\omega) = \varepsilon_{00}(\omega)$.  Substituting (eqs~\ref{chi0} and \ref{matrixelements}) into (eq~\ref{epsilonmatrix}), and suppressing $k$-point dependence, we obtain the simplified form
\begin{equation}
\Imag[\varepsilon(\omega)] = \frac{4\pi\eta}{\Omega} \sum_{nm}\sum_{ss'} \frac{f_{ms} - f_{ns'}}{(\omega - \varepsilon_{ns'} + \varepsilon_{ms})^2 + \eta^2}
 \left(\frac{\hat{e}_{\textbf{q}}\cdot\langle\psi_{ns'}|{\mathbf{\nabla}} |\psi_{ms}\rangle}{\varepsilon_{ns'} - \varepsilon_{ms}}\right)^2.\label{Imeps}
\end{equation}
where $\hat{e}_{\textbf{q}}$ is a unit vector parallel to $\textbf{q}$.  
The matrix elements in (eq~\ref{Imeps}) may be expressed as
\begin{eqnarray}
\langle\psi_{ns'}|{\mathbf{\nabla}} |\psi_{ms}\rangle &=& \sum_{\mu\nu} c_{\nu ns'}^*c_{\mu ms}^{} \Big[\langle \widetilde{\phi}_{\nu}|{\mathbf{\nabla}}|\widetilde{\phi}_{\mu}\rangle +\nonumber\\
&& \sum_{aij} P_{ins'}^{a*}\left[\langle\phi_i^a|\mathbf{\nabla}|\phi_j^a\rangle - \langle\widetilde{\phi}_i^a|\mathbf{\nabla}|\widetilde{\phi}_j^a\rangle\right] P_{jms}^a\Big] ,
\end{eqnarray}
where $\widetilde{\phi}_{\nu}$ and $\phi_{\nu}$ are the pseudo and all-electron locally centered atomic orbitals (LCAOs) within the projector augmented wave (PAW) formalism with coefficients $c_{\nu n s'}$ for the $n$th KS wave function in spin channel $s'$, and PAW projectors
\begin{equation}
P^{a*}_{ins'} = \langle\widetilde{\psi}_{ns'}|\widetilde{p}_{i}^a\rangle,
\end{equation}
of the $n$th pseudo KS wave function in spin channel $s'$ $\widetilde{\psi}_{ns'}$ onto the $i$th LCAO orbital of atom $a$.  

\begin{figure}[!h]
\noindent{\color{StyleColor}{\rule{\columnwidth}{1pt}}}
\includegraphics[scale=0.42]{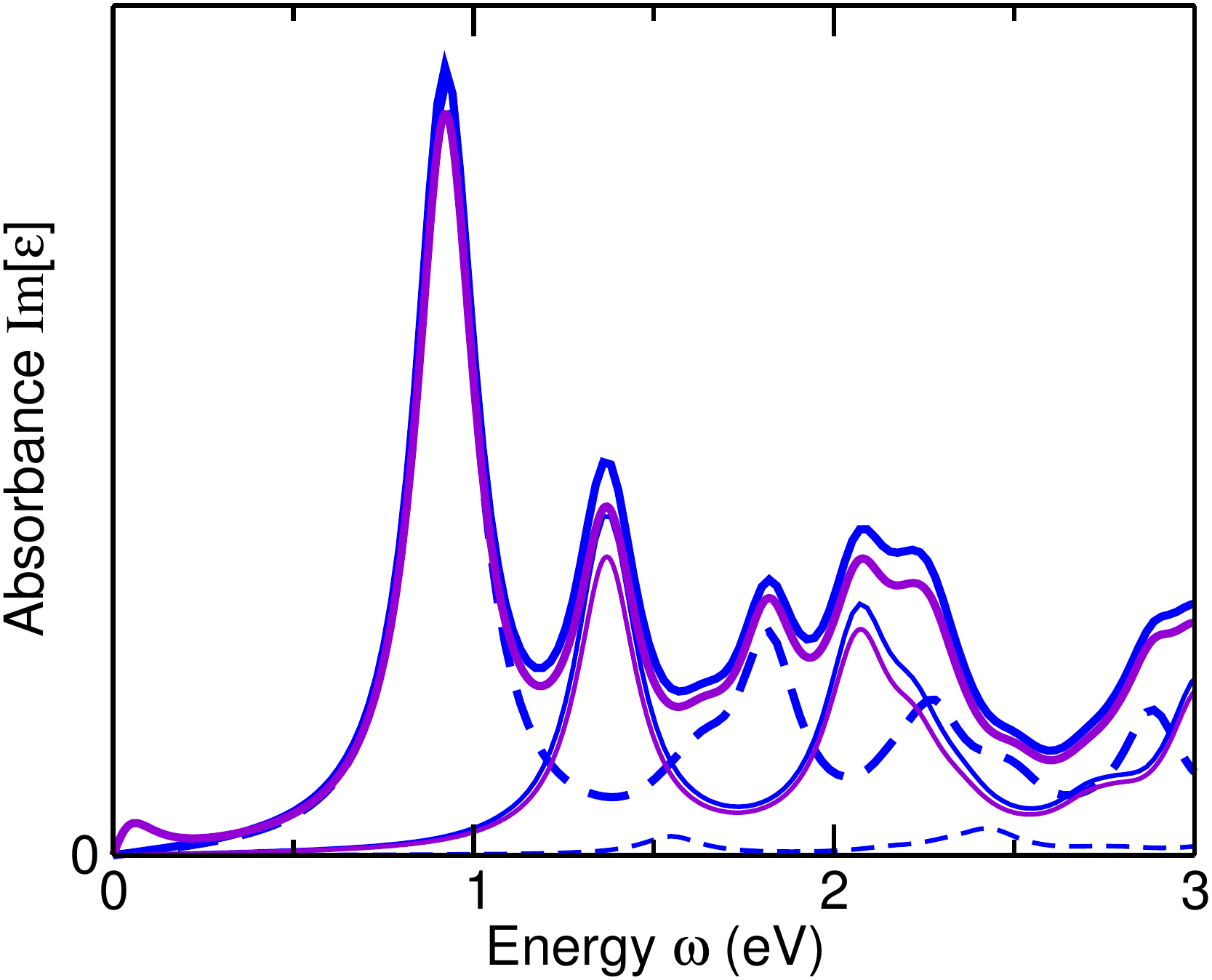}
\caption{TDDFT-RPA absorbance $\Imag[\varepsilon]$ of a (6,5) SWNT in the ground state versus energy in eV obtained from our LCAO implementation (violet) and a real space projection of the wave functions (blue) including (dashed) and neglecting (solid) local field effects.  Total absorbance (thick lines) and absorbance for light polarized perpendicular to the SWNT axis (thin lines) are shown.
}\label{figS1}
\noindent{\color{StyleColor}{\rule{\columnwidth}{1pt}}}
\end{figure}

In Figure~\ref{figS1}
 we compare the TDDFT-RPA spectra obtained within the LCAO basis (violet), and after projecting onto a real space grid (blue). The latter was calculated using the TDDFT-RPA implementation within the \textsc{gpaw} code \cite{GPAW,GPAWRev,GPAWLCAO} described in refs \citenum{response1}, \citenum{response2} and \citenum{DuncanGrapheneTDDFTRPA}\nocite{response1,response2,DuncanGrapheneTDDFTRPA} including 79 \textbf{G} vectors. 

When local field effects (LFEs) are neglected, both codes agree up to a constant of proportionality. We attribute this difference to the projection of the LCAO basis functions onto the real space grid. When including LFEs, the absorbance perpendicular to the SWNT axis is greatly suppressed, while that along the axis is unchanged. However, this suppression of the perpendicular absorbance may be due to modeling debundled SWNTs\cite{SWCNTbundleRPA}.  In any case, we find the neglect of LFEs is justified for these systems, especially because our primary interest is the $E_{11}$ transition along the SWNT axis.

\section*{
\large$\blacksquare$\normalsize\ 
ASSOCIATED CONTENT}

\subsubsection*{
\includegraphics[height=8pt]{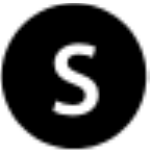} 
Supporting Information
}
\noindent
The Supporting Information is available free of charge on the \href{http://pubs.acs.org}{ACS Publications website} at DOI: \href{http://dx.doi.org/10.1021/acs.jpcc.5b10025}{10.1021/acs.jpcc.5b10025}. 

LCAO TDDFT-RPA source code (\href{http://pubs.acs.org/doi/suppl/10.1021/acs.jpcc.5b10025/suppl_file/jp5b10025_si_001.zip}{ZIP})

\section*{
\large$\blacksquare$\normalsize\ 
AUTHOR INFORMATION}
\subsubsection*{Corresponding Author}
\noindent 
*E-mail: \href{mailto:duncan.mowbray@gmail.com}{duncan.mowbray@gmail.com}.
\subsubsection*{Notes} 
\noindent The authors declare no competing financial interest.
\section*{
\large$\blacksquare$\normalsize\ 
ACKNOWLEDGMENTS} 

We acknowledge financial support from the European Projects POCAONTAS (FP7-PEOPLE-2012-ITN No.~316633), DYNamo (ERC-2010-AdG-267374), OLIMPIA (P7-PEOPLE-212-ITN No.~316832); Spanish Grants (FIS20113-46159-C3-1-P); Italian Grants (GGP12033); Grupos Consolidados UPV/EHU del Gobierno Vasco (IT-578-13); and computational time from the BSC Red Espanola de Supercomputacion.

\bibliography{bibliography}


\end{document}